\documentclass[iicol,pdflatex,sn-basic]{sn-jnl}

\usepackage{multirow}
\usepackage{makecell}
\usepackage{graphicx}
 \usepackage{booktabs}
\usepackage{amsmath}

\usepackage{tcolorbox}
\usepackage{enumitem}

\usepackage{graphicx}%
\usepackage{multirow}%
\usepackage{amsmath,amssymb,amsfonts}%
\usepackage{amsthm}%
\usepackage{mathrsfs}%
\usepackage[title]{appendix}%
\usepackage{xcolor}%
\usepackage{textcomp}%
\usepackage{manyfoot}%
\usepackage{booktabs}%
\usepackage{algpseudocode}%
\usepackage{listings}%
\usepackage{multirow}
\usepackage{makecell}
\usepackage{graphicx}
 \usepackage{booktabs}
\usepackage{amsmath}
\newcounter{algorithm}

\theoremstyle{thmstyleone}%
\theoremstyle{thmstyletwo}%

\theoremstyle{thmstylethree}%

\begin{document}

\title[Article Title]{Coordinated Incentives in AI-Generated Misinformation Governance}


\author[1]{\fnm{Qin} \sur{Li}}

\author[2]{\fnm{Gui} \sur{Zhang}}

\author*[2]{\fnm{Minyu} \sur{Feng}}\email{myfeng@swu.edu.cn}

\author*[3,4,5,6]{\fnm{Matja{\v z}} \sur{Perc}}\email{matjaz.perc@gmail.com}

\author[7]{\fnm{Attila} \sur{Szolnoki}}

\affil[1]{%
  \orgdiv{Business College}, 
  \orgname{Southwest University}, 
  \orgaddress{%
    \city{Chongqing}, 
    \postcode{402460}, 
    \country{China}
  }
}

\affil[2]{%
  \orgdiv{College of Artificial Intelligence}, 
  \orgname{Southwest University}, 
  \orgaddress{%
    \city{Chongqing}, 
    \postcode{400715}, 
    \country{China}
  }
}

\affil[3]{%
  \orgdiv{Faculty of Natural Sciences and Mathematics}, 
  \orgname{University of Maribor}, 
  \orgaddress{%
    \city{Maribor}, 
    \postcode{2000}, 
    \country{Slovenia}
  }
}

\affil[4]{%
  \orgname{Community Healthcare Center Dr. Adolf Drolc Maribor}, 
  \orgaddress{%
    \city{Maribor}, 
    \postcode{2000}, 
    \country{Slovenia}
  }
}

\affil[5]{%
  \orgdiv{University College}, 
  \orgname{Korea University}, 
  \orgaddress{%
    \city{Seoul}, 
    \postcode{02841}, 
    \country{Republic of Korea}
  }
}

\affil[6]{%
  \orgdiv{Department of Physics}, 
  \orgname{Kyung Hee University}, 
  \orgaddress{%
    \city{Seoul}, 
    \postcode{02447}, 
    \country{Republic of Korea}
  }
}

\affil[7]{%
  \orgdiv{Institute of Technical Physics and Materials Science}, 
  \orgname{Centre for Energy Research}, 
  \orgaddress{%
    \postcode{1525}, 
    \city{Budapest}, 
    \country{Hungary}
  }
}

\abstract{With the rapid diffusion of AI-generated content, AI-driven misinformation is becoming increasingly pervasive and difficult to govern, undermining information credibility and social trust. This study models the strategic interdependence among a government regulator, an AI enterprise, and users through a three-party evolutionary game that incorporates heterogeneous rewards and punishments. From the resulting replicator equations, we characterize the evolutionary stability of competing governance and production strategies. The analysis indicates that neither unilateral regulation nor market incentives alone can effectively curb misinformation. Instead, an evolutionarily stable regime of real-information production arises only when regulatory rewards and punishment intensity, enterprise reputation loss, and user adoption incentives collectively surpass critical thresholds. The findings highlight the need for coordinated and adaptive policy mixes that align regulatory instruments with enterprise behavior and user uptake while managing governance costs.}


\maketitle

\section{Introduction}

In recent years, the rapid development of Artificial Intelligence Generated Content (AIGC) has profoundly changed the way information is produced and disseminated (\cite{chen2024empowering,xu2024unleashing}). AI systems, represented by large language models and multimodal generative models, can generate text (\cite{radford2021learning}), images (\cite{vaswani2017attention}), and video (\cite{vondrick2016generating}) at extremely low cost and high efficiency, showing an exponential diffusion trend on social media platforms (\cite{liu2025optimizing}). This technological breakthrough has significantly improved content production capabilities and facilitated access to and diversification of information (\cite{sui2025will}). However, it also 
lowered the threshold for generating false information, making the automated and large-scale production of misleading content a reality, thus posing a serious challenge to the information ecosystem and social trust structure (\cite{zhang2024detecting}).

Through a systematic interdisciplinary literature review from 1991 to the present, (\cite{sanfilippo2025sociotechnical}) develop a misinformation governance model, propose a future research agenda, and offer recommendations for contextually adaptive and comprehensive governance. However, compared to traditional dissemination of misinformation, AI-driven misinformation exhibits greater concealment, realism, and speed of dissemination (\cite{davis2025disinformation, xu2023combating}). On the one hand, generative models can simulate authoritative discourse styles and highly realistic visual content, significantly improving the credibility of information  (\cite{zhu2023storytrans}); On the other hand, the viral spread of AI content on social networks makes regulation significantly lag behind 
adequate regulation is unable to follow the speed of spread of AI content 
(\cite{park2025generative}). Furthermore, the costs of platform review and manual verification are constantly rising, and there is a significant 
that is accompanied by an  
efficiency imbalance between government regulation and technological governance (\cite{cajueiro2025comprehensive}). Therefore, 
how to characterize the formation and spread mechanisms of AI-generated misinformation in a complex environment of multi-stakeholder interaction, and to construct reasonable governance strategies, has become an important research issue in the field of information security and intelligent governance (\cite{zhou2023synthetic}). 

Given the prominent conflicts of interest and strategic interdependency among AI enterprises, users, and government regulators in information generation and dissemination, this issue is inherently a typical social dilemma that can be modeled by a multi-agent game system (\cite{wang2022cooperative}). Traditional static game theory fails to capture the dynamic evolutionary characteristics of strategic choices (\cite{wang1989static}), whereas Evolutionary Game Theory(EGT) enables the dynamic representation of strategy adjustment of 
boundedly 
rational agents during long-term interactions, thus offering an effective analytical tool to investigate behavioral evolution in misinformation governance (\cite{weibull1997evolutionary,sigmund1999evolutionary}). Recently, this approach has been widely applied to problems in biology (\cite{leimar2023game,mcnamara2020game}), economics (\cite{archetti2011economic,kabir2020evolutionary}), and society (\cite{akccay2020deconstructing, jusup2022social}), which has been shown to be suitable for analyzing the policy stability of complex social systems.

In EGT, the replicator dynamic, first proposed by Taylor and Jonker, is one of the core methods to characterize the nonlinear interest relationships between different stakeholders (\cite{taylor1978evolutionary}). Since then, this method has been widely used in various three-party evolutionary games. For example, \cite{li2023open} proposed a three-party evolutionary game model involving service providers, users, and regulators. Using replicator dynamics and numerical simulations, their study analyzed how factors such as implementation costs, data value, regulatory incentives, and mining capacity influence cooperative behavior among stakeholders. Furthermore, this technique has been applied in mobile crowdsourcing by modeling the behaviors of different stakeholders, and
with results 
showing that reward and punishment mechanism boost trustworthy participation and effectively curb free-riding and false reporting (\cite{li2021three}). Based on existing research, 
reward and punishment mechanism (\cite{zhang2025spatial,zhang2025evolution})
in EGT 
offers a proper solution to the cooperation dilemma among stakeholders that has been widely studied in several studies
(\cite{motepalli2021reward,wu2017probabilistic,zu2022reward}).

However, existing studies on misinformation governance largely rely on simplified two-party or traditional three-party evolutionary game models, where reward and punishment mechanisms are often idealized and lack alignment with real-world complexity. In particular, such frameworks fail to adequately capture the interdependent interactions among government regulators, AI enterprises, and users, and tend to overlook heterogeneous incentives and nonlinear dynamics. Compared with these approaches, our study incorporates heterogeneous reward–punishment mechanisms and models the tripartite interactions in a more realistic and dynamically coupled manner.

To address these limitations, this paper develops a three-party evolutionary game model that integrates government regulators, AI enterprises, and users within a unified analytical framework. By embedding heterogeneous incentive schemes and nonlinear interaction structures, the proposed model systematically characterizes the strategic evolution of each stakeholder under AI-generated misinformation governance. This approach not only enriches the theoretical modeling of multi-agent evolutionary dynamics but also provides deeper insights into the design of effective governance mechanisms, encompassing regulatory policy formulation, enterprise-level strategic responses, and user adoption behaviors. In sum, the key contributions of our work are as follows:

\begin{itemize}
	\item This system integrates government regulators, AI information generation enterprises, and users into a unified dynamic framework based on EGT. It reveals the root causes of the formation and spread of AI-generated misinformation from a complex, multi-stakeholder perspective.
	\item We develop a reward and punishment payoff function that systematically captures the practical incentive asymmetry and risk transmission mechanisms from a multi-agent perspective.
	\item We construct the replicator dynamics system from the payoff matrix, identify the key factors and their coupling effects shaping AI enterprises' information generation, user adoption, and government regulation decisions, and derive the evolutionary stable strategy (ESS) in governance scenarios using Jacobian local stability analysis, theoretically characterizing the long-term behavioral patterns induced by diverse institutional environments.
    \item Both theory and simulation show that relying solely on punishment and reward is insufficient to suppress misinformation and that there is a significant synergistic threshold effect between regulatory rewards, punishment intensity, and reputation loss.
\end{itemize}

The rest of this paper is structured as follows. In Section~\ref{model}, we establish a three-party evolutionary game among government regulators, AI enterprises, and users with a heterogeneous reward and punishment mechanism and theoretically deduce and analyze the ESS of this replicator dynamics system. In Section~\ref{simulation}, we verified our theoretical analysis by simulation. Our key findings are summarized in Section~\ref{Conclusion} which also provides some recommended policies to avoid the negative consequences of misinformation.

\section{An evolutionary game model in AI information generation}\label{model}
\begin{figure*}
    \centering
    \includegraphics[width=1\linewidth]{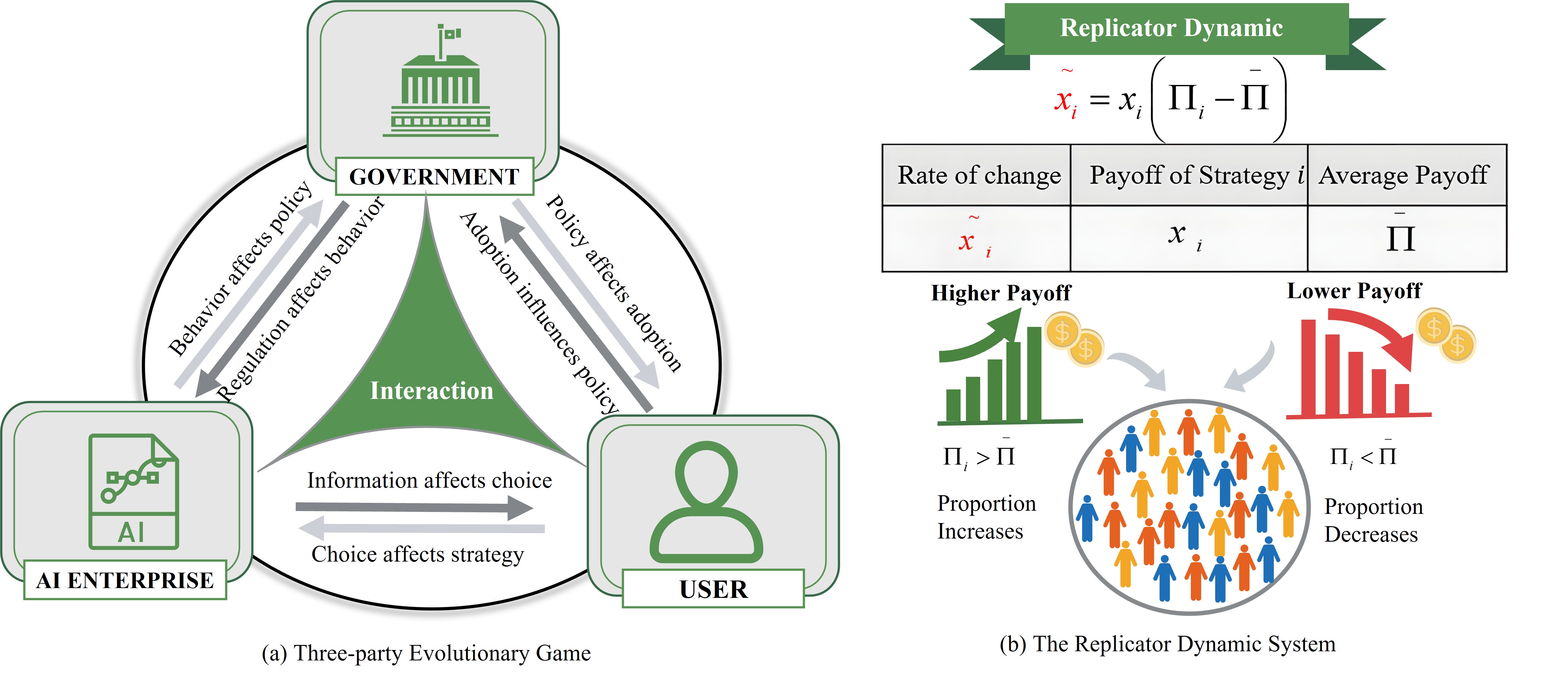}
    \caption{\textbf{Diagram of the three-party evolutionary game model including government regulators, AI enterprises, and users.} (a) The nonlinear interaction among government regulators, AI enterprises, and users. (b) Diagram of a replicator dynamic system. }
    \label{diag}
\end{figure*}

In the midst of the proliferation of AI-generated information, the risk of algorithms creating or amplifying misinformation has become increasingly prominent, severely undermining social trust and mounting pressures on public governance. Therefore, effectively modeling AI misinformation governance by means of EGT is an essential theoretical tool to address this problem. Based on this research path, we construct a three-party evolutionary game model involving government regulators, AI enterprises, and users with heterogeneous rewards and punishments. More specifically, we make the fundamental assumption that government regulators opt for two alternative strategic choices, namely regulation and non-regulation, with the corresponding adoption ratios of \(z\) and \(1-z\), respectively. At the same time, users select between two optional strategies of adopting information and not adopting information with the respective implementation ratios of \(y\) and \(1-y\). AI enterprises employ two distinct operational approaches of generating real information and generating false information with the respective application ratios of \(x\) and \(1-x\) in the tripartite evolutionary game system, and $x,y,z\in(0,1)$. Based on the three stakeholders, we propose an EGT model to study the competition and cooperation among government regulators, users, and AI enterprises, and further investigate the ESS of these three participants. Fig.~\ref{diag} illustrates the game-theoretic relationship among government regulators, users, and AI enterprises regarding the governance of misinformation and a schematic diagram of the model using replication dynamics to study the changes in the proportion of participants.

In the following subsection, we establish the three-party evolutionary game based on the assumption, which incorporates payoff calculation, the establishment of a replicator dynamic system, and conditions for the equilibrium point to become the ESS to provide effective suggestions for combating misinformation.
\subsection{The model setup}
Here, we have established a three-party evolutionary game involving government regulators, AI enterprises, and users, in which all are finite and rational groups. In this three-party game, each party has two strategies, through which they interact with the other two parties and gain their own payoff. Due to the inherent characteristics of replication dynamics, the strategy adoption ratio of a participant increases when their individual payoff exceeds the average payoff of the entire group and declines in the opposite scenario, as illustrated in Fig.~\ref{diag}(b). 

The main goal of all three groups is to maximize their own payoff when interacting with others. Therefore, the definition of payoff functions for each stakeholder in the tripartite evolutionary game constitutes a core determinant that exerts a pivotal influence on the directional evolution of the game system, and against the backdrop of extant literature and practical scenarios, we thus put forward the following research assumptions for subsequent analysis:

\begin{enumerate}
   \item \textbf{For the government regulator}, the regulation generates a supervisory benefit $B_g$ that reflects social governance gains. When users adopt real information under regulation, the government also receives a market return $R_2$. However, regulation incurs an administrative cost of $C_3$. If the government does not regulate, it suffers a governance loss $D_g$ due to the diffusion of misinformation and the accompanying erosion of public trust. Accordingly, the payoff to government regulators is determined by the trade-off between the regulatory benefits $(B_g, R_2)$ and regulatory costs $C_3$ incurred under regulatory strategies and the systemic losses $D_g$ arising from the adoption of non-regulatory strategies.
   \item \textbf{For the AI enterprise}, generating real information requires operational cost $C_1$, but yields a market return $R_1$ when users adopt it. Under government regulation, compliant enterprises receive a reward of $F$. Enterprise generating false information face reputation damage $L_1$ due to loss of credibility. In the presence of government regulation, such entities incur an additional punishment of $\alpha F$, where the parameter $\alpha$ denotes the intensity of the punitive measures imposed. In the absence of regulation, enterprises avoid direct punishment but still incur reputation loss if users adopt false information. Hence, enterprise payoff reflects the balance among compliance reward, production cost, potential punishment, and reputation risks.
   \item \textbf{For the user}, adoption has an informational value $V_1$ when the information is accurate, but causes harm $V_2$ when it is false. Users always incur an adoption cost of $C_2$ when choosing to rely on AI information. If users do not adopt AI information, they suffer a loss $L_2$ due to missing potentially useful content. Therefore, the user's payoff depends on perceived information reliability, adoption cost, and the trade-off between informational gain and avoidance of misinformation risk.
\end{enumerate}

All parameters are assumed to be non-negative, ensuring consistent interpretations as benefits, costs, or losses. Government-related parameters $(B_g, R_2, C_3, D_g)$ capture regulatory gains and risks, enterprise parameters $(R_1, C_1, F, \alpha, L_1)$ reflect market returns, incentives, and punishment, and user parameters $(V_1, V_2, C_2, L_2)$ describe the trade-off between information value and risk. These settings guarantee a coherent payoff structure for evolutionary analysis. The notations adopted in this model are briefly summarized in Table~\ref{t1}.
\begin{table*}[h]
    \centering
    \caption{\centering{Notation Table}}
    \label{t1}
    \resizebox{\textwidth}{!}{
    \begin{tabular}{@{}cc@{}}
        \toprule[1.5pt]
        \makebox[0.2\textwidth][c]{\textbf{Symbol}} &
        \makebox[0.8\textwidth][c]{\textbf{Definition}} \\
        \midrule[1pt]

        $x$ & Fraction of AI enterprise choosing to generate real information \\

        $y$ & Fraction of users choosing to adopt information provided by AI enterprise \\

        $z$ & Fraction of government regulator choosing to regulate AI enterprise \\

        $B_g$ & Government benefit obtained from regulating AI enterprise\\

        $D_g$ & Loss of government caused by non-regulation \\

        $F$ & Reward granted to compliant AI enterprise under regulation\\

        $\alpha$ & Punishment intensity coefficient ($\alpha F$ represents regulatory punishment) \\

        $R_1$ & Market return gained by AI enterprise when real information is adopted by the user\\

        $R_2$ & Market return gained by the government when real information is adopted by the user \\

        $C_1$ & Cost for AI enterprise to generate real information \\

        $C_2$ & Cost incurred by user when adopting AI-generated information \\

        $C_3$ & Cost of government regulation \\

        $V_1$ & Value for user  when adopting real information \\

        $V_2$ & Loss for user when adopting false information \\

        $L_1$ & Loss to AI enterprises caused by generating false information \\

        $L_2$ & Loss from not adopting potentially useful AI information \\

        \bottomrule[1.5pt]
    \end{tabular}
    }
\end{table*}

\begin{table*}[htbp]
\centering
\caption{\centering{Payoff Matrix}}
\label{tab:payoff_matrix}
\renewcommand{\arraystretch}{1.5}
\resizebox{\textwidth}{!}{
\begin{tabular}{|c|c|c|c|}
\hline
\multirow{3}{*}{Government regulator} &
\multirow{3}{*}{AI enterprise} &
\multicolumn{2}{c|}{User} \\ 

& & Adopt $(y)$ & Not adopt $(1-y)$ \\ \hline

\multirow{2}{*}{Regulate $(z)$}
& Real information $(x)$
& $B_g+R_2-C_3,\; R_1+F-C_1,\; V_1-C_2$
& $B_g-C_3,\; F-C_1,\; -L_2$ \\ 

& False information $(1-x)$
& $B_g-C_3,\; -\alpha F-L_1,\; -V_2-C_2$
& $B_g-C_3,\; -\alpha F,\; -L_2$ \\ \hline

\multirow{2}{*}{Not regulate $(1-z)$}
& Real information $(x)$
& $-D_g,\; R_1-C_1,\; V_1-C_2$
& $-D_g,\; -C_1,\; -L_2$ \\ 

& False information $(1-x)$
& $-D_g,\; -L_1,\; -V_2-C_2$
& $-D_g,\; 0,\; -L_2$ \\ \hline

\end{tabular}
}
\end{table*}

\subsection{Payoff calculation for the three stakeholders}
Based on the assumptions and notations mentioned above, we construct a payoff matrix for the three stakeholders in line with realistic scenarios in Table~\ref{tab:payoff_matrix}. As an illustration, the payoff $\Pi_{A_1}$ of the AI company generating real information can be described as
\begin{equation}
\begin{aligned}
\Pi_{A_1} =& yz(R_1+F-C_1) + z(1-y)(F-C_1) \\
       &+ y(1-z)(R_1-C_1) + (1-y)(1-z)(-C_1)\\
      =&yR_1 + zF - C_1\,.
\end{aligned}
\end{equation}
Here, the payoff $\Pi_{A1}$ can be decomposed into four parts which have the following meanings:

\begin{enumerate}
    \item \textit{Users adopt information under regulation}: The term $yz (R_1 + F - C_1)$ represents the case in which the government enforces the regulation and users adopt the generated information, so the AI enterprise obtains a market return $R_1$ from successful adoption, receives a regulatory reward $F$ for compliance, and incurs the production cost $C_1$, with the payoff weighted by the joint probability $yz$.
    \item \textit{Users do not adopt information under regulation}: The term $z(1-y)(F - C_1)$ represents the case in which the government enforces the regulation but users do not adopt the generated information. As a result, no market revenue is generated, while the AI enterprises still receive regulatory incentive $F$ and incurs the production cost $C_1$, with the payoff weighted by the joint probability $z(1-y)$.
    \item \textit{Users adopt information without regulation}:
    The term $y(1-z)(R_1 - C_1)$ represents the case in which the government does not enforce the regulation, but users adopt the generated information. Accordingly, the AI enterprise gains only the market return $R_1$ without receiving any regulatory reward, while still incurring the production cost $C_1$, with the payoff weighted by the joint probability $y(1-z)$.
    \item \textit{Users do not adopt information without regulation}:
    The term $(1-y)(1-z)(-C_1)$ captures the worst-case scenario in which the government does not enforce the regulation, and users do not adopt the generated information. Consequently, the AI enterprises gains no positive revenue and only incurs the production cost $C_1$, with the payoff weighted by the joint probability $(1-y)(1-z)$.
\end{enumerate}

Similarly, the payoff of AI enterprises that generate false information under government regulation can be expressed as follows,

\begin{equation}
\begin{aligned}
\Pi_{A_2} &= zy(-\alpha F - L_1) + z(1-y)(-\alpha F)\\
         &\quad + (1-z)y(-L_1) + (1-z)(1-y)\cdot 0 \\
         &= -z\alpha F - yL_1,
\end{aligned}
\end{equation}
considering the two types of payoffs, $\Pi_{A_1}$ and $\Pi_{A_2}$ corresponding to AI enterprises' production of real information and false information, respectively, we can further calculate the expected payoff $\Pi_A$ for AI enterprises,
\begin{equation}
\begin{aligned}
\Pi_A &= x\Pi_{A_1} + (1-x)\Pi_{A_2} \\
      &= x(yR_1 + zF - C_1) + (1-x)(-z\alpha F - yL_1).
\end{aligned}
\end{equation}

Similarly, the payoff values for users who adopt or ignore AI-generated information are defined as $\Pi_{U_1}$ and $\Pi_{U_2}$ which is calculated as
\begin{equation}
    \begin{cases}
        \Pi_{U_1} =  xV_1 - (1-x)V_2 - C_2,\ \\
        \Pi_{U_2} =  -L_2\,,
    \end{cases}
\label{Eq: 2}
\end{equation}
we get the expected payoff for users is 
\begin{equation}
\begin{aligned}
\Pi_U &= y\Pi_{U_1} + (1-y)\Pi_{U_2}. \\
      &= y\big[xV_1 - (1-x)V_2 - C_2\big]+(1-y)(-L_2).
\end{aligned}
\end{equation}

Last, the payoff values of government regulators who choose to regulate or not regulate AI enterprises are denoted by $\Pi_{R_1}$ and $\Pi_{R_2}$, respectively. These payoffs can be derived using the same analytical framework as above, by accounting for the corresponding strategy profiles and payoff structures. Accordingly, $\Pi_{R_1}$ and $\Pi_{R_2}$ are calculated as

\begin{equation}
    \begin{cases}
        \Pi_{R_1} = xyR_2 + B_g - C_3, \\
        \Pi_{R_2} = -D_g,
    \end{cases}
\label{Eq:}
\end{equation}
therefore, the expected payoff of this stakeholder is
\begin{equation}
\begin{aligned}
\Pi_R &= z\Pi_{R_1} + (1-z)\Pi_{R_2}. \\
      &= z(xyR_2 + B_g - C_3) +(1-z)(-D_g).
\end{aligned}
\end{equation}

In summary, this section calculated the payoff for the three parties under different strategy combinations. The next subsection will then derive replication dynamics

\subsection{The replicator dynamics system}
Drawing on the foundational studies of evolutionary dynamics by Friedman (\cite{friedman_d_e91}) and Taylor and Jonker (\cite{taylor1978evolutionary}), strategy evolution in a population can be described by continuous-time differential equations. In this framework, a strategy expands its presence whenever its expected return exceeds the mean payoff of the population, while strategies with inferior performance gradually diminish. More precisely, the instantaneous change in the proportion of individuals adopting a given strategy is determined by the difference between its own payoff and the population-wide average payoff. Therefore, the replication dynamic equations describing the interaction among government regulators, AI enterprises, and users can be expressed as follows,
\begin{equation}
\left\{
\begin{aligned}
\dot{x}
&= x(\Pi_{A_1}-\Pi_A) \\
&= x(1-x)\underbrace{\left[ y(R_1+L_1) + zF(1+\alpha) - C_1 \right]}_{\Pi_{A_1}-\Pi_{A_2}}, \\
\dot{y}
&= y(\Pi_{U_1}-\Pi_U) \\
&= y(1-y)\underbrace{\left[ x(V_1+V_2) - V_2 - C_2 + L_2 \right]}_{\Pi_{U_1}-\Pi_{U_2}}, \\
\dot{z}
&= z(\Pi_{R_1}-\Pi_R) \\
&= z(1-z)\underbrace{\left[ xyR_2 + B_g + D_g - C_3 \right]}_{\Pi_{R_1}-\Pi_{R_2}}.
\end{aligned}
\right.
\label{Eq:replication_dynamics_with_underbrace}
\end{equation}

Eq.~\eqref{Eq:replication_dynamics_with_underbrace} constitutes the continuous frequency dynamic system for the tripartite evolutionary game involving AI enterprises, users, and government regulators. Through the above stepwise derivation, we have obtained the replicator dynamic equations of the three stakeholder populations. 
These equations reveal that the proportion of each stakeholder choosing a specific strategy is a time-varying variable that evolves with the game process. In addition, the magnitude and sign of the right-hand side of each equation reflect the speed and direction of the change in the proportion of the corresponding population’s strategy selection over time, laying a theoretical foundation for the subsequent stability analysis of the evolutionary game system.

\subsection{Stability analysis}
After establishing the replication dynamics equations for government regulators, AI enterprises, and users, it is essential to conduct stability analysis under different parameter settings. The ESS of the differential equation system can be derived through local stability analysis of the system's Jacobian matrix. Based on Eq.~\eqref{Eq:replication_dynamics_with_underbrace}, we can obtain the Jacobian matrix of this three-party evolutionary game model.
Let ($f_x=\dot{x}$), ($f_y=\dot{y}$), and ($f_z=\dot{z}$). The Jacobian matrix $J(x,y,z)$ is defined as
\begin{equation}
J(x,y,z)=
\begin{pmatrix}
\dfrac{\partial f_x}{\partial x} & \dfrac{\partial f_x}{\partial y} & \dfrac{\partial f_x}{\partial z} \\[6pt]
\dfrac{\partial f_y}{\partial x} & \dfrac{\partial f_y}{\partial y} & \dfrac{\partial f_y}{\partial z} \\[6pt]
\dfrac{\partial f_z}{\partial x} & \dfrac{\partial f_z}{\partial y} & \dfrac{\partial f_z}{\partial z}
\end{pmatrix},\
\end{equation}
by explicitly differentiating the replicator dynamics in Eq.~\eqref{Eq:replication_dynamics_with_underbrace}, the corresponding partial derivatives are obtained as

\begin{equation}
\left\{
\begin{aligned}
\dfrac{\partial f_x}{\partial x} &=(1-2x)\bigl[y(R_1+L_1)+zF(1+\alpha)-C_1\bigr], \\
\dfrac{\partial f_x}{\partial y} &=x(1-x)(R_1+L_1), \\
\dfrac{\partial f_x}{\partial z} &=x(1-x)F(1+\alpha), \\
\dfrac{\partial f_y}{\partial x} &=y(1-y)(V_1+V_2), \\
\dfrac{\partial f_y}{\partial y} &=(1-2y)\bigl[x(V_1+V_2)-V_2-C_2+L_2\bigr], \\
\dfrac{\partial f_y}{\partial z} &=0, \\
\dfrac{\partial f_z}{\partial x} &=z(1-z)yR_2, \\
\dfrac{\partial f_z}{\partial y} &=z(1-z)xR_2, \\
\dfrac{\partial f_z}{\partial z} &=(1-2z)\bigl[xyR_2+B_g+D_g-C_3\bigr].
\end{aligned}
\right.
\end{equation}
\begin{table*}[h]
  \centering
  \caption{\centering Eigenvalues and ESS conditions  in the three-party evolutionary game}

  \label{tab:ess_eigenvalues_conditions}

  \resizebox{\textwidth}{!}{

  \begin{tabular}{ccccc}

    \toprule

    ESS & Eigenvalue $\lambda_1$ & Eigenvalue $\lambda_2$ & Eigenvalue $\lambda_3$ & ESS Conditions \\

    \midrule

    $E_1(0,0,0)$ & $-C_1$ & $L_2-V_2-C_2$ & $B_g+D_g-C_3$ & $\begin{cases} V_2+C_2>L_2 \\ B_g+D_g<C_3 \end{cases}$ \\

    $E_2(0,0,1)$ & $F(1+\alpha)-C_1$ & $L_2-V_2-C_2$ & $C_3-B_g-D_g-R_2$ & $\begin{cases} F(1+\alpha)<C_1 \\ V_2+C_2>L_2 \\ B_g+D_g+R_2>C_3 \end{cases}$ \\

    $E_3(0,1,0)$ & $L_1+R_1-C_1$ & $C_2+V_2-L_2$ & $B_g+D_g-C_3$ & $\begin{cases} L_1+R_1<C_1 \\ C_2+V_2<L_2 \\ B_g+D_g<C_3 \end{cases}$ \\

    $E_4(0,1,1)$ & $F(1+\alpha)+L_1+R_1-C_1$ & $C_2+V_2-L_2$ & $C_3-B_g-D_g$ & $\begin{cases} F(1+\alpha)+L_1+R_1<C_1 \\ C_2+V_2<L_2 \\ B_g+D_g>C_3 \end{cases}$ \\

    $E_5(1,0,0)$ & $C_1$ & $L_2+V_1-C_2$ & $B_g+D_g-C_3$ & $\begin{cases} C_1<0 \\ V_1+L_2<C_2 \\ B_g+D_g<C_3 \end{cases}$ \\

    $E_6(1,0,1)$ & $C_1-F(1+\alpha)$ & $L_2+V_1-C_2$ & $C_3-B_g-D_g$ & $\begin{cases} F(1+\alpha)>C_1 \\ L_2+V_1<C_2 \\ B_g+D_g>C_3 \end{cases}$ \\

    $E_7(1,1,0)$ & $C_1-L_1-R_1$ & $C_2-L_2-V_1$ & $B_g+D_g+R_2-C_3$ & $\begin{cases} L_1+R_1>C_1 \\ L_2+V_1>C_2 \\ B_g+D_g+R_2<C_3 \end{cases}$ \\

    $E_8(1,1,1)$ & $C_1-F(1+\alpha)-L_1-R_1$ & $C_2-L_2-V_1$ & $C_3-B_g-D_g-R_2$ & $\begin{cases} F(1+\alpha)+L_1+R_1>C_1 \\ L_2+V_1>C_2 \\ B_g+D_g+R_2>C_3 \end{cases}$ \\

    \bottomrule

  \end{tabular}}

\end{table*}
The equilibrium points can be obtained by setting $\dot x=\dot y=\dot z=0$ in Eq.~\eqref{Eq:replication_dynamics_with_underbrace}. This system has eight pure strategy equilibrium points, including $E_1(0,0,0)$, $E_2(0,0,1)$, $E_3(0,1,0)$, $E_4(0,1,1)$, $E_5(1,0,0)$, $E_6(1,0,1)$, $E_7(1,1,0)$ and $E_8(1,1,1)$. Furthermore, we can obtain the asymptotically stable interior equilibrium point as $(x, y, z) = (x^{\star}, y^{\star}, z^{\star})$, which satisfies

\begin{equation}
\begin{cases}
y^{\star}(R_1+L_1) + z^{\star}F(1+\alpha) - C_1 = 0, \\
x^{\star}(V_1+V_2) - V_2 - C_2 + L_2 = 0, \\
x^{\star}y^{\star}R_2 + B_g + D_g - C_3 = 0,
\end{cases}
\end{equation}
where $0<x^{\star}<1$, $0<y^{\star}<1$, and $0<z^{\star}< 1$. This system of equations defines the interior equilibrium point \((x^{\star}, y^{\star}, z^{\star})\) of the three-party replicator dynamic system. Based on this, we can further obtain the trace of the Jacobian matrix in this case, which is calculated as
\begin{equation}
\begin{aligned}
tr(J) 
=&\ \dfrac{\partial f_x}{\partial x}
+\dfrac{\partial f_y}{\partial y}
+\dfrac{\partial f_z}{\partial z} \\
=&\ (1-2x)\bigl[y(R_1+L_1)+zF(1+\alpha)-C_1\bigr] \\
+&(1-2y)\bigl[x(V_1+V_2)-V_2-C_2+L_2\bigr] \\
+&(1-2z)\bigl[xyR_2+B_g+D_g-C_3\bigr]\\
=&0.
\end{aligned}
\end{equation}

Since the trace of the matrix is $tr(J)=\lambda_{1}+\lambda_{2}+\lambda_{3}$, not all eigenvalues can be negative simultaneously. Therefore, the internal equilibrium point \((x^{\star}, y^{\star}, z^{\star})\)  is excluded from the discussion of stability analysis. We focus on pure strategy equilibrium points, taking $(x,y,z)=(1,1,1)$ as an example, we derive the conditions for it to become an ESS. In this case, the element of the Jacobian matrix takes the form given by the following equation:

\begin{equation}
\left\{
\begin{aligned}
&\dfrac{\partial f_x}{\partial x} = C_1 - y(R_1+L_1) - zF(1+\alpha), \\
&\dfrac{\partial f_y}{\partial y} = V_2 + C_2 - x(V_1+V_2) - L_2, \\
&\dfrac{\partial f_z}{\partial z} = C_3 - xyR_2 - B_g - D_g, \\
&\dfrac{\partial f_x}{\partial y},\dfrac{\partial f_x}{\partial z},\dfrac{\partial f_y}{\partial x},\dfrac{\partial f_y}{\partial z},\dfrac{\partial f_z}{\partial x},\dfrac{\partial f_z}{\partial y}=0.
\end{aligned}
\right.
\label{eq:12}
\end{equation}

In this case, the Jacobian matrix degenerates into a diagonal matrix, from which the eigenvalues of the Jacobian matrix in this equilibrium can be obtained as ($\lambda_1=C_1 - y(R_1+L_1) - zF(1+\alpha), \lambda_2=V_2 + C_2 - x(V_1+V_2) - L_2, \lambda_3=C_3 - xyR_2 - B_g - D_g$). Based on evolutionary game theory, an equilibrium point in the replicator dynamics system is asymptotically stable if and only if all eigenvalues of the corresponding Jacobian matrix have negative real parts, i.e., $\lambda_1<0$, $\lambda_2<0$, $\lambda_3<0$. We can thus derive the stability conditions for the equilibrium point $(x, y, z) = (1, 1, 1)$, which is given by

\begin{equation}
\begin{cases}
F(1+\alpha)+L_1+R_1>C_1, \\
L_2+V_1>C_2, \\
B_g+D_g+R_2>C_3.
\end{cases}
\label{condition(1,1,1)}
\end{equation}

Eq.~\eqref{condition(1,1,1)} characterizes the evolutionary stability condition of the pure-strategy equilibrium $\mathrm{ESS},(1,1,1)$. This condition ensures that, once the system state enters a neighborhood of $(1,1,1)$, the evolutionary trajectories of the tripartite game will asymptotically converge to this equilibrium under the replicator dynamics, provided that the model parameters satisfy the specified inequalities. In this sense, $\mathrm{ESS}(1,1,1)$ represents a locally asymptotically stable state of the system, reflecting the robustness of the corresponding strategy profile against small perturbations. Furthermore, by employing the same analytical framework, we systematically derive the local stability conditions for the remaining seven equilibrium points. For clarity and ease of comparison, the eigenvalue structures and the corresponding evolutionary stability conditions of all equilibria are summarized in Table~\ref{tab:ess_eigenvalues_conditions}.

\section{Simulation results}\label{simulation}
To verify the theoretical predictions obtained in Section~\ref{model}, we perform numerical simulations. All numerical experiments and visualizations in this study were conducted using Python 3.11.7. The computational framework is built on the scientific computing ecosystem, with core dependencies on NumPy for array manipulation and numerical operations, SymPy for symbolic derivation and expression construction, SciPy for solving ordinary differential equations via odeint, and Matplotlib for scientific plotting and result visualization. 

\refstepcounter{algorithm}
\begin{tcolorbox}[
  title={Algorithm \thealgorithm: Simulation of Replicator Dynamics},
  colback=white,
  colframe=black,
  boxrule=0.5pt,
  sharp corners
]

\textbf{Input:} Replicator dynamic system 
$\dot{\mathbf{s}} = F(\mathbf{s},\Theta)$; 
parameter set $\Theta = \{\theta_1,\theta_2,\dots\}$; 
scanning subsets $\Theta_i$; 
initial state $\mathbf{s}_0$; 
time horizon $T$.

\medskip
\textbf{Output:} Equilibrium state $\mathbf{s}^*$ under different parameter configurations.

\begin{enumerate}[leftmargin=1.4em]
\item Define payoff functions and construct the corresponding replicator dynamic equations.
\item Initialize data structures to store equilibrium outcomes.
\item For each parameter subset $\Theta_i$ under investigation:
  \begin{enumerate}
  \item For each parameter combination in $\Theta_i$:
    \begin{enumerate}
    \item Substitute the current parameters into the dynamic system.
    \item Numerically integrate $\dot{\mathbf{s}} = F(\mathbf{s},\Theta)$ over $[0,T]$.
    \item Obtain the terminal state $\mathbf{s}(T)$.
    \item Treat $\mathbf{s}(T)$ as an approximation of the equilibrium state $\mathbf{s}^*$.
    \item Store $\mathbf{s}^*$ corresponding to the current parameter configuration.
    \end{enumerate}
  \end{enumerate}
\item Return the parameter equilibrium mapping $\Theta_i \mapsto \mathbf{s}^*$.
\end{enumerate}

\label{alg:generic_replicator_sensitivity}
\end{tcolorbox}

The simulation approach is presented in Algorithm~\ref{alg:generic_replicator_sensitivity}, which provides a systematic computational framework for analyzing replicator dynamics under varying parameter configurations. Specifically, the algorithm begins by formulating the payoff structure and the corresponding replicator dynamic system. It then performs a structured parameter scanning process over predefined subsets of the parameter space.
For each parameter configuration, the dynamic system is instantiated and numerically integrated from a given initial state over a finite time horizon. The resulting terminal state is regarded as an approximation of the equilibrium outcome. By iterating this process across all parameter combinations, the algorithm constructs a mapping from parameter subsets to their associated equilibrium states. This procedure enables a comprehensive sensitivity analysis of the evolutionary outcomes with respect to model parameters, thereby revealing how variations in the underlying payoff structure influence the stability and stationary behavior of the system.

\subsection{Evolutionary trajectories of stakeholders}
\begin{figure*}
    \centering
    \includegraphics[width=1\linewidth]{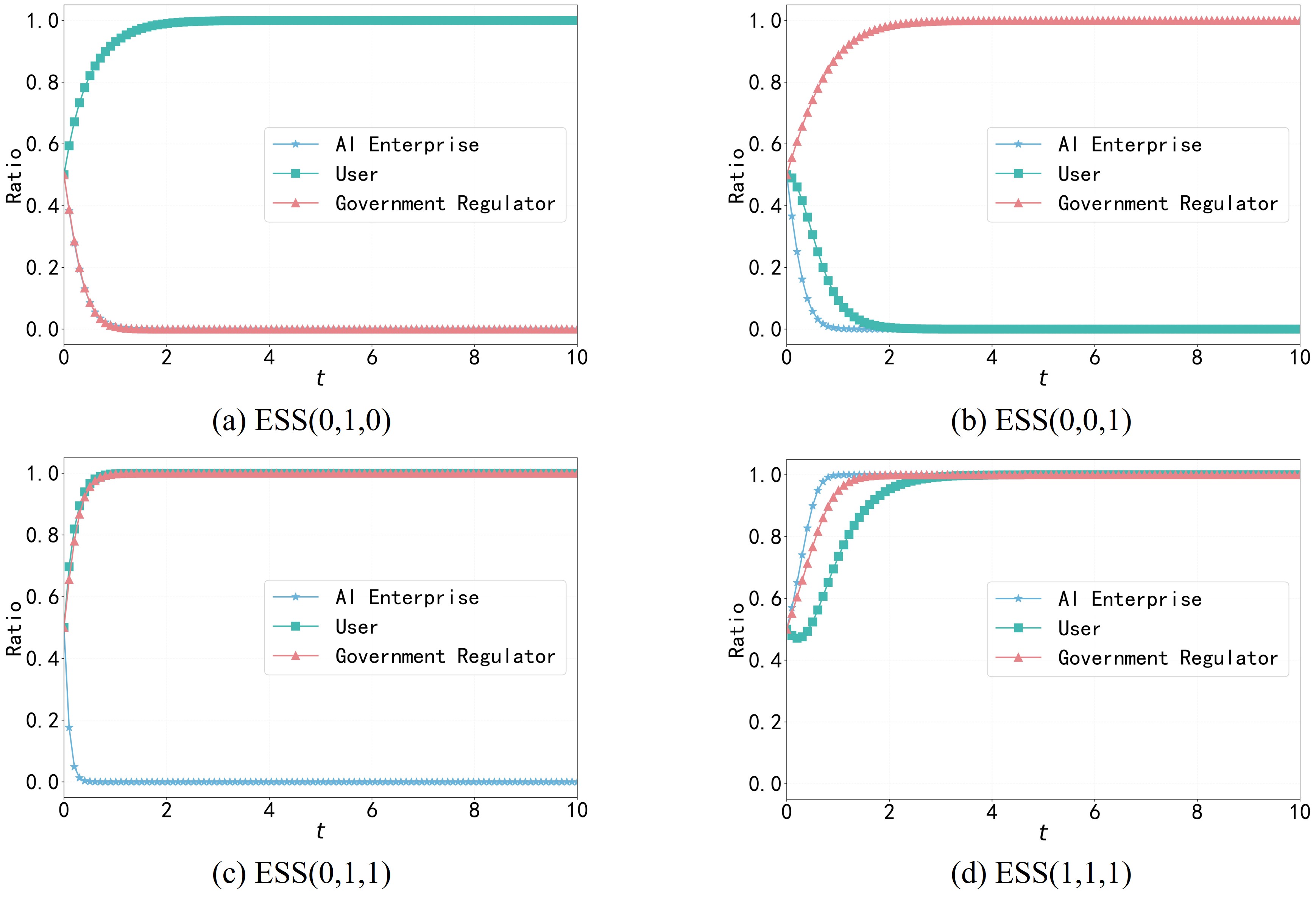}
    \caption{\textbf{The evolutionary trajectories of AI enterprises, users, and government regulators.}  The $x$ and $y$ axes are set as the ratio of player and time step and the simulation is initialized at the state $(x,y,z)=(0.5,0.5,0.5)$. (a) $B_g=2$, $D_g=1$, $F=3$, $\alpha=0.5$, $R_1=3$, $R_2=2$, $C_1=10$, $C_2=4$, $C_3=8$, $V_1=2$, $V_2=2$, $L_1=3$, and $L_2=8$ to satisfy the condition for (0,1,0) to become ESS. (b) $B_g=6$, $D_g=5$, $F=2$, $\alpha=0.5$, $R_1=3$, $R_2=1$, $C_1=10$, $C_2=5$, $C_3=9$, $V_1=2$, $V_2=4$, $L_1=3$, and $L_2=6$   to satisfy the condition for (0,0,1) to become ESS. (c)$B_g=10$, $D_g=8$, $F=2$, $\alpha=0.5$, $R_1=3$, $R_2=2$, $C_1=20$, $C_2=5$, $C_3=12$, $V_1=3$, $V_2=4$, $L_1=2$, and $L_2=15$ to satisfy the condition for (0,1,1) to become ESS. (d) $B_g=6$, $D_g=5$, $F=10$, $\alpha=0.5$, $R_1=5$, $R_2=4$, $C_1=10$, $C_2=8$, $C_3=10$, $V_1=5$, $V_2=1$, $L_1=5$, and $L_2=5$  to satisfy the condition for (1,1,1) to become ESS. }
    \label{fig:placeholder}
\end{figure*}

Based on Fig.~\ref{fig:placeholder} and the theoretical analysis of the stability of the pure strategy equilibrium points above, this subsection verifies the evolution results of the system under different parameter conditions through numerical simulation. The focus is on examining whether the system can converge from the general initial state to the theoretically derived ESS when the corresponding inequality conditions are met. All simulations are conducted on the basis of the replicator dynamics equations. For given parameter combinations and identical initial conditions, the temporal evolution trajectories of the strategy adoption ratios of AI enterprises, users, and government regulators are tracked to verify the validity of the stability analysis of our theory.

\begin{figure*}[h]
    \centering
    \includegraphics[width=1\linewidth]{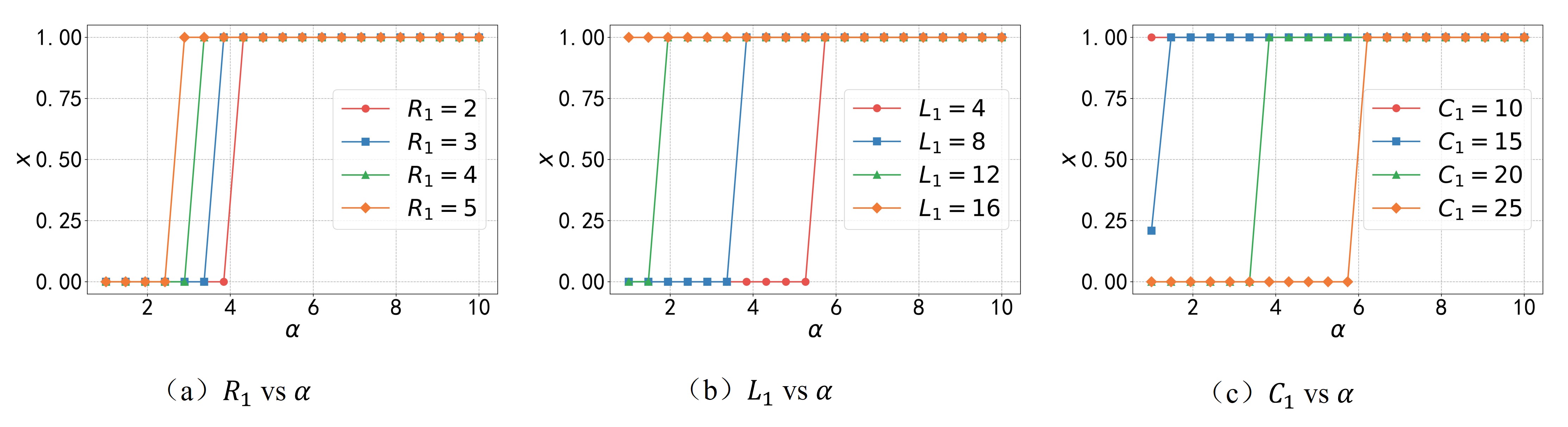}
    \caption{\textbf{Influence of $R_1$, $L_1$, $C_1$, and $\alpha$ on the AI enterprises to generate information.} All panels adopt a unified parameter configuration: $B_g=10$, $D_g=8$, $F=2$, $R_2=2$, $V_1=3$, $V_2=4$, $C_2=5$, $C_3=12$, $L_2=15$. The simulation is initialized at the state $(x,y,z)=(0.5,0.5,0.5)$ with a time horizon of $t\in[0,10]$.
    (a) $L_1=8$, $C_1=20$, and $R_1=[2,3,4,5]$. (b) $R_1=3$, $C_1=20$, and $L_1=[4,8,12,16]$. (c) $R_1=3$, $L_1=8$, and $C_1=[10,15,20,25]$. The $x$ and $y$ axes are set as the punishment intensity coefficient and
    the fraction of AI enterprises generating true information, respectively.}
    \label{3}
\end{figure*}
\begin{enumerate}
    \item \textit{Trajectories of stakeholders at ESS(0, 1, 0):} When \(L_1+R_1<C_1\), \(C_2+V_2<L_2\) and \(B_g+D_g<C_3\) hold, the local stability analysis reveals that the strategy combination of \((0,1,0)\) constitutes the ESS. Under such parameter conditions, AI enterprises tend to refrain from generating real information, users opt to adopt AI-generated information, and government regulators are inclined to abstain from regulatory interventions. The parameters are set as $B_g=2$, $D_g=1$, $F=3$, $\alpha=0.5$, $R_1=3$, $R_2=2$, $C_1=10$, $C_2=4$, $C_3=8$, $V_1=2$, $V_2=2$, $L_1=3$, and $L_2=8$ to satisfy the condition for (0,1,0) to become ESS which is shown in Fig.~\ref{fig:placeholder}(a). Notably, the proportion of AI enterprises opting to generate real information gradually converges to 1, whereas the proportions of users choosing to adopt AI-generated information and government regulators electing to implement regulatory measures for AI enterprises decline to 0 along nearly identical evolutionary trajectories.
    \item \textit{Trajectories of stakeholders at ESS(0, 0, 1):} When \(F(1+\alpha)<C_1\), \(C_2+V_2>L_2\) and \(B_g+D_g+R_2>C_3\) hold, the equilibrium point \((0,0,1)\) satisfies the ESS condition. In this case, AI enterprises opt not to generate real information, users refrain from adopting AI-generated information, and government regulators proactively implement regulatory strategies. The parameters are set as $B_g=6$, $D_g=5$, $F=2$, $\alpha=0.5$, $R_1=3$, $R_2=1$, $C_1=10$, $C_2=5$, $C_3=9$, $V_1=2$, $V_2=4$, $L_1=3$, and $L_2=6$   to satisfy the condition for (0,0,1) to become ESS shown in Fig.~\ref{fig:placeholder}(b). Here, the ratio of government regulators implementing oversight on AI enterprises converges to 1 over time, whereas the users adopting AI-generated information and that of AI enterprises producing real information both decline to 0. In particular, the latter exhibits a more rapid rate of decline.
    
    \item \textit{Trajectories of stakeholders at ESS(0, 1, 1):} When \(F(1+\alpha)+R_1+L_1<C_1\), \(C_2+V_2<L_2\) and \(B_g+D_g>C_3\) hold, the equilibrium point \((0,1,1)\) is identified as the ESS. This implies that AI enterprises refrain from generating real information, users opt to adopt AI-generated information, and government regulators persist in implementing regulatory oversight.  The parameters are set as $B_g=10$, $D_g=8$, $F=2$, $\alpha=0.5$, $R_1=3$, $R_2=2$, $C_1=20$, $C_2=5$, $C_3=12$, $V_1=3$, $V_2=4$, $L_1=2$, and $L_2=15$ to satisfy the condition for (0,1,1) to become ESS. As shown in \ref{fig:placeholder}(c), the proportion of government regulators adopting regulatory strategies and that of users opting to adopt AI-generated information converges steadily to 1, whereas the ratio of AI enterprises producing real information declines consistently to 0.
    
    \item \textit{Trajectories of stakeholders at ESS(1, 1, 1):} When \(F(1+\alpha)+R_1+L_1>C_1\), \(L_2+V_1>C_2\) and \(B_g+D_g+R_2>C_3\) hold, the equilibrium point \((1,1,1)\) meets all evolutionary stability criteria and thus qualifies as the ESS. Parameters are set as \(B_g=6\), \(D_g=5\), \(F=10\), \(\alpha=0.5\), \(R_1=5\), \(R_2=4\), \(C_1=10\), \(C_2=8\), \(C_3=10\), \(V_1=5\), \(V_2=1\), \(L_1=5\) and \(L_2=5\), which satisfies the parameter conditions for the equilibrium point \((1,1,1)\) to be the ESS. In Fig.~\ref{fig:placeholder}(d), the proportions of government regulators implementing oversight on AI enterprises, users adopting AI-generated information, and AI enterprises producing real information all converge gradually to 1. AI enterprises exhibit the fastest rate of increase, followed by government regulators and then users.
\end{enumerate}

By analyzing the evolution trajectories under different parameters, we verified the ESS under different conditions in Table~\ref{tab:ess_eigenvalues_conditions}, which are consistent with the theoretical results obtained in Section~\ref{model}.

\begin{figure*}
    \centering
    \includegraphics[width=1\linewidth]{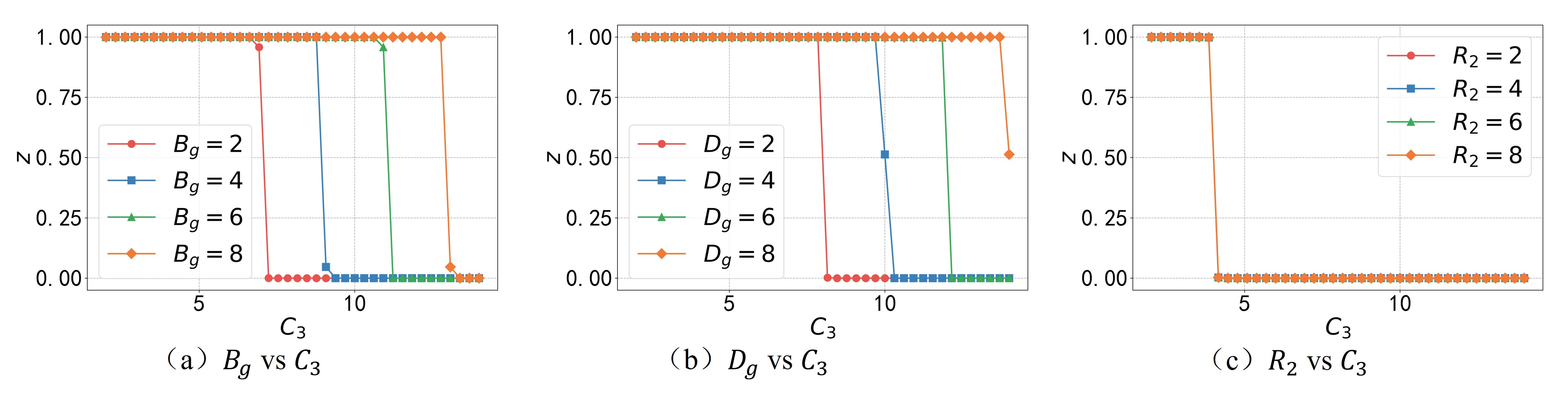}
    \caption{\textbf{Impact of $R_2$, $B_g$, $D_g$ and $C_3$ on government regulators}. All panels are simulated under a unified baseline parameter setting: $F=2$, $\alpha=0.5$, $R_1=3$, $C_1=10$, $C_2=5$, $V_1=2$, $V_2=4$, $L_1=3$, $L_2=6$, with initial state $(x,y,z)=(0.5,0.5,0.5)$ and time horizon $t\in[0,10]$. (a) $D_g=5$, $R_2=1$, and $B_g={2,4,6,8}$. (b) $B_g=6$, $R_2=1$, and $D_g={2,4,6,8}$. (c) $B_g=2$, $D_g=2$, and $R_2={2,4,6,8}$. The $x$ and $y$ axes are set as the cost of government regulation and the fraction of the government regulating AI enterprise, respectively.}
    \label{4}
\end{figure*}
\begin{figure*}
    \centering
    \includegraphics[width=1\linewidth]{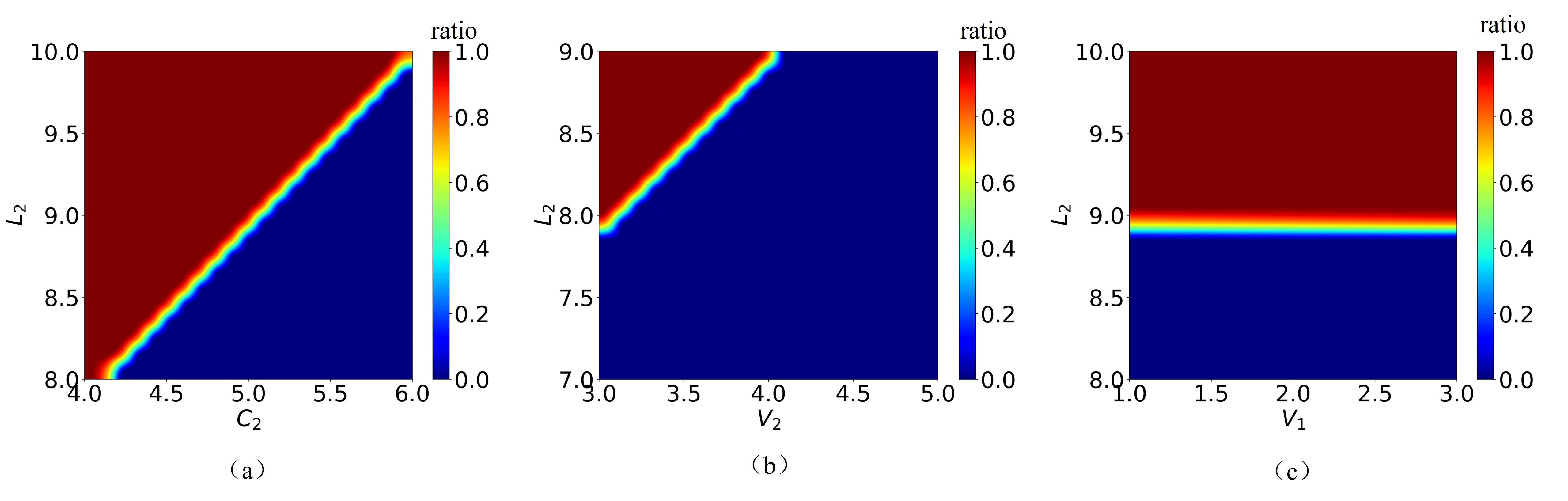}
    \caption{\textbf{Heat maps of the stationary fraction of users adopting information generated by the AI enterprise under different parameter combinations}. Initial condition: $(x_0,y_0,z_0)=(0.5,0.5,0.5)$  with a time step size $t=10$. Fixed parameters: $B_g=6$, $D_g=5$, $F=2$, $\alpha=0.5$, $R_1=3$, $R_2=1$, $C_1=10$, $C_3=9$, and $L_1=3$. 
    (a) $C_2 \in [4,6]$, $L_2 \in [8,10]$, $V_1=2$, and $V_2=4$.
    (b) $V_2 \in [3,5]$, $L_2 \in [7,9]$, $C_2=5$, and $V_1=2$.
    (c) $V_1 \in [1,3]$, $L_2 \in [8,10]$, $V_2=4$, and $C_2=5$.
    The simulation step size for all varying parameters is $0.01$. }
    \label{fig:5} %
\end{figure*}
\subsection{Stationary strategies of AI enterprises}
This subsection describes the evolutionary dynamic of strategy selection by AI enterprises with variations in key payoff-related parameters \(R_1\), \(L_1\), \(C_1\), and \(\alpha\). All simulations are implemented under a unified baseline parameter setting, with the initial state specified as \((x,y,z)=(0.5,0.5,0.5)\) and time horizon $t\in[0,10]$, thereby isolating the marginal effects of each individual parameter on the evolutionary trajectory of \(x\). The punishment intensity coefficient $\alpha$ is set in the range of $[1,10]$.

As illustrated in Fig.~\ref{3}(a), it depicts the impact of market returns \(R_1\) with fixed values of \(L_1\) and \(C_1\). We set $R_1$ in the range $[2, 5]$. As \(R_1\) increases, the growth rate of \(x\) rises markedly and the stationary proportion of AI enterprises producing real information increases simultaneously. When \(\alpha\) is sufficiently large, the system converges to a high \(x\) equilibrium, indicating that stronger market incentives effectively drive AI enterprises to adopt real information generation strategies.
When we explore the impact of reputation loss $L_1$ caused by AI enterprise generating misinformation on their strategies in Fig.~\ref{3}(b) with $L_1$ in the range of $[4,16]$, the larger the reputation damage incurred by AI enterprises from producing false information, the more inclined they are to opt for the strategy of generating real information. Furthermore, the higher the cost for AI enterprises to produce real information, the more inclined they are to generate false information, as depicted in \ref{3}(c) with $C_1$ in the range of $[10,25]$. When the production cost of real information for AI enterprises is low (\(C_1=10\)), their evolved stationary strategy is to produce real information, irrespective of the magnitude of the punishment factor \(\alpha\).

Overall, Fig.~\ref{3} demonstrates that the strategic evolution of AI enterprises is jointly driven by economic incentives, reputation loss, production cost, and regulatory punishment intensity. These numerical results are fully consistent with the analytical results of the replicator dynamics in Eq.~\eqref{Eq:replication_dynamics_with_underbrace}, and they highlight the necessity of coordinating reward, punishment, and cost to induce real information generation.

\subsection{Stationary strategies of regulators}
This subsection depicts the evolutionary dynamic of strategy selection by the government regulator with variations in key payoff-related parameters \(B_g\), \(D_g\), \(R_2\), and \(C_3\). The initial strategy configuration of the stakeholder is $(x, y, z) = (0.5, 0.5, 0.5)$ and the evolution step size is $t = 10$. We mainly study the change in the ratio of regulatory cost $C_3$ with the range $[2,14]$ for government regulators' choice of regulation under different conditions $R_1$, $L_1$, and $C_1$.

The simulation results are presented in Fig.~\ref{4} . As the regulatory cost $C_3$ rises, the government regulators tend to adopt an inaction strategy, that is, they may refrain from overseeing the conduct of AI enterprises. As shown in Fig.~\ref{4}(a), an increase in regulatory benefits \(B_g\) significantly elevates the proportion of government regulators adopting regulatory strategies, indicating that higher direct governance returns effectively boost regulators’ incentives to implement oversight. Furthermore, the greater the reputation loss \(D_g\) incurred by regulators from failing to regulate AI enterprises, the more inclined they are to adopt a regulatory strategy. This reflects regulators' aversion to incurring substantial losses by pursuing a non-regulatory strategy. Interestingly, the proportion of government regulators choosing to regulate AI enterprises does not change as the market return $R_2$ increases, illustrated in Fig.~\ref{4}(c). When the regulatory cost \(C_3\) of government regulators is relatively low, the parameter combination converges to \(E_2(0, 0, 1)\); as \(C_3\) rises to a certain threshold, the combination shifts to \(E_1(0, 0, 0)\). In both scenarios, with users refraining from adopting information and AI enterprises failing to generate real information, the government’s market return \(R_2\) is attenuated by the product of \(x\) and \(y\), as defined in Eq.~\eqref{Eq:replication_dynamics_with_underbrace}. Accordingly, the government’s regulatory ratio \(z\) shows an identical trend at varying levels of market return \(R_2\). Overall, these results suggest that regulatory benefits, loss avoidance, and social returns jointly shape the evolutionary dynamics of government regulation, driving the system toward a stable regulatory outcome in line with the stability conditions of theory.

\subsection{Stationary strategies of users}
We now examine the stationary adoption behavior of users by jointly varying the key parameters, including $C_2$, $L_2$, $V_1$, and $V_2$ that define their payoff function. Specifically, heat maps are presented to illustrate the equilibrium fraction of users adopting AI-generated information under different combinations of adoption cost, information value, misinformation loss, and non-adoption loss, while keeping other parameters fixed, where the initial state is set to $(x, y, z) = (0.5, 0.5, 0.5)$ with a time horizon $t \in [0,10]$, the baseline parameters are fixed as $B_g = 6$, $D_g = 5$, $F = 2$, $\alpha = 0.5$, $R_1 = 3$, $R_2 = 1$, $C_1 = 10$, $C_3 = 9$, and $L_1 = 3$, while the adoption cost $C_2$, information value $V_1$, misinformation loss $V_2$, and non-adoption loss $L_2$ are varied pairwise within the specified ranges to examine their coupled effects on users’ evolutionary adoption behavior.

\begin{figure*}
    \centering
    \includegraphics[width=1\linewidth]{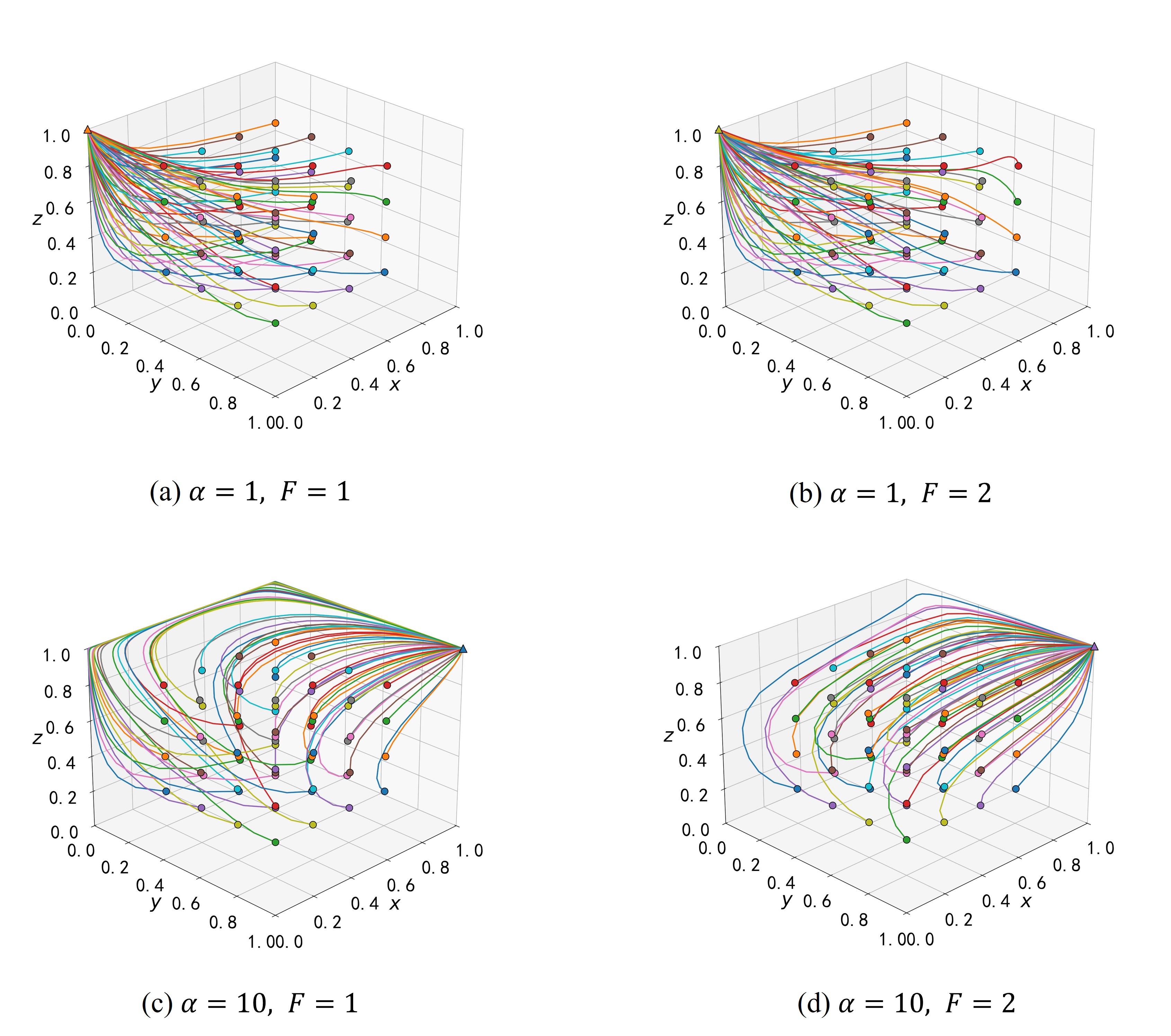}
    \caption{\textbf{Evolutionary trajectories under different initial configurations}. For AI enterprises, government regulators, and users, the initial strategy configuration is set within the range of $[0.2, 0.8]$ with a step size of 0.2, leading to a total of 64 initial strategy configurations. The evolution step size is $t = 30$ and circles represent the starting point of evolution, and triangles represent the ending point in the simulation. Fixed parameters: $B_g=6$, $D_g=5$, $F=2$, $R_1=3$, $R_2=1$, $C_1=10$, $C_2=5$,$C_3=9$, $V_1=2$, $V_2=4$, $L_1=3$ and $L_2=6$. 
    (a) $\alpha=1,F=2$.
    (b)  $\alpha=10,F=2$.
    (c)  $\alpha=1,F=1$.
    (d) $\alpha=10,F=1$.
    }
    \label{fig:6} %
\end{figure*}
Fig. ~\ref{fig:5} displays the heat maps of the stationary ratio of users that adopt the information generated by the AI enterprise for three representative parameter settings. These three distinct parameter combinations collectively demonstrate that a higher $L_2$ loss incurred by users who forego AI-generated information corresponds to a larger proportion of users adopting information. Specifically, the higher the user cost \(C_2\) and the greater the negative impact \(V_2\) of erroneous information generated by AI enterprise, the more inclined users are to choose the strategy of not using AI-generated information, as shown in Fig.~\ref{fig:5}(a) and in Fig.~\ref{fig:5}(b). Interestingly, in the parameter configuration of Fig.~\ref{fig:5}(c), the proportion of users adopting AI-generated information does not rise with the positive benefits \(V_1\) from AI enterprise’s real information. The system converges to two stationary states \(E_2(0,0,1)\) and \(E_4(0,1,1)\) determined by \(L_2\), meaning AI enterprises do not generate authentic information. Thus, due to the replicator dynamics equation in Eq.~\eqref{Eq:replication_dynamics_with_underbrace}, \(V_1\) has no effect on users’ adoption strategies when an AI enterprise chooses to generate false information.

\subsection{Evolutionary trajectories under different initial configurations}
To further examine the robustness of evolutionary outcomes and exclude the possibility that convergence behavior observed in previous simulations is driven by specific initial conditions, we investigate the evolutionary trajectories of the three-party system under multiple initial strategy configurations. In this simulation, the initial strategy proportions of AI enterprises, users, and government regulators are independently varied within the interval $[0.2, 0.8]$ with a step size of 0.2, resulting in 64 distinct initial states. For each initial configuration, the replicator dynamic system is numerically integrated over a time horizon of $t = 30$. The baseline parameters are fixed as $B_g = 6, D_g = 5, F = 2, R_1 = 3, R_2 = 1, C_1 = 10, C_2 = 5, C_3 = 9, V_1 = 2, V_2 = 4, L_1 = 3$, and $L_2 = 6$. 

To explore the combined effects of regulatory punishment intensity and reward strength, four representative scenarios are considered by varying the punishment intensity coefficient $\alpha$ and the regulatory reward $F$, as illustrated in Fig.~\ref{fig:6}. When both the reward and punishment coefficients are relatively small,  $(\alpha = 1, F = 1)$ in Fig.~\ref{fig:6}(a) and $(\alpha = 1, F = 2)$ in Fig.~\ref{fig:6}(b), the replicator dynamic system satisfies the stability condition $F(1+\alpha)<C_1,  V_2+C_2>L_2 $, and $B_g+D_g+R_2>C_3$ of $E_2(0,0,1)$ listed in Table~\ref{tab:ess_eigenvalues_conditions}. Consequently, regardless of the initial strategy configuration, the evolutionary trajectories of the system converge toward $E_2$, indicating that weak regulatory incentives lead to a robust non-cooperative outcome. In contrast, we increased the reward and punishment coefficients, as shown in Fig.~\ref{fig:6}(c)($\alpha=10, F=1$) and Fig.~\ref{fig:6}(d)($\alpha=10, F=2$). The three-party evolution reaches the ideal $E_8(1,1,1)$ state for the high reward and severe punishment coefficients satisfy ($F(1+\alpha)+L_1+R_1>C_1$, $L_2+V_1>C_2$, and $ B_g+D_g+R_2>C_3$ ) in Table~\ref{tab:ess_eigenvalues_conditions}. In particular, high reward and punishment coefficients enable the system to enter ESS more quickly. Only by balancing these two coefficients allow the system achieve the ideal state($E_8(1,1,1)$) where governments regulate AI enterprises, AI enterprises generate real information, and users effectively adopt the information.

\section{Conclusion and policy recommendation}\label{Conclusion}
\subsection{Strengths and weaknesses}
Our principal motivation is to reveal how AI enterprises that potentially produce misinformation may modify the interactions of governmental and individual actors in social dilemma situations. To address this issue, we construct a three-party evolutionary game model involving government regulators, AI enterprises, and users, and investigate the governance mechanisms of AI-generated misinformation under heterogeneous reward and punishment schemes. By deriving the replicator dynamic system and conducting local stability analysis, we identify the ESS under different parameter configurations and validate the theoretical findings through numerical simulations. The results reveal pronounced strategic interdependence and nonlinear feedback among the three stakeholders.

Nevertheless, this study has some limitations. The model is developed under the assumption of a well-mixed population and simplified reward and punishment mechanisms, which may not fully capture the heterogeneity of real-world interactions and the adaptiveness of practical governance processes. Future research can extend this framework to structured populations, such as complex networks and heterogeneous interaction environments, and incorporate more realistic, adaptive mechanisms or data-driven approaches to better reflect practical misinformation governance scenarios.

\subsection{Policy Recommendations}

The analysis further demonstrates that relying solely on regulatory punishment or market incentives is insufficient to suppress AI-generated misinformation in the long run. Effective governance emerges only when regulatory rewards, punishment intensity, reputation loss, and user adoption incentives jointly exceed certain critical threshold conditions. In particular, a socially desirable equilibrium—where AI enterprises generate truthful information, users actively adopt credible content, and governments maintain effective oversight—can be achieved only through coordinated policy design that balances regulatory costs and social returns. These findings provide a theoretical foundation for developing systematic and sustainable governance policies for AI-generated misinformation. Based on the analytical and simulation results, the following policy recommendations are proposed:

\begin{itemize}
  \item \textbf{Establish balanced reward and punishment mechanisms for AI enterprises.}  
    Regulatory authorities should design incentive schemes combining appropriate rewards for compliant behavior with sufficiently strong penalties for misinformation generation. Our results indicate that rewards alone cannot offset production costs, while punishment without incentives may discourage compliance. A calibrated combination of both mechanisms is necessary to induce AI enterprises to consistently generate real information.
  \item \textbf{Reduce regulatory cost while enhancing governance return.}  
 Governments should improve regulatory efficiency by adopting technological tools including automated auditing, algorithmic transparency mechanisms, and AI-assisted supervision. Lower regulatory costs and higher social returns significantly expand the parameter region where active regulation is evolutionarily stable.
  \item \textbf{Strengthen user-side incentives and risk awareness.}  
  Policies should aim to increase the perceived value of accurate information and reduce users’ adoption costs through information labeling, credibility indicators, and public education. Enhancing users’ ability to distinguish misinformation shifts evolutionary dynamics toward equilibria with higher information quality.

  \item \textbf{Leverage reputation mechanisms as long-term governance instruments.}  
In addition to formal regulation, transparent reputation systems for AI enterprises should be established. Reputation loss acts as a persistent constraint even in low-regulation environments, reinforcing compliance incentives, and discouraging opportunistic misinformation behavior.
  \item \textbf{Adopt adaptive and scenario dependent regulatory strategies.}  
Given multiple ESS, uniform regulatory policies may be ineffective. Regulatory intensity and incentive structures should be dynamically adjusted according to market conditions, misinformation risk levels, and user behavioral responses.

\end{itemize}

Overall, these policy recommendations highlight the necessity of a coordinated, multi-dimensional governance framework that aligns incentives across governments, AI enterprises, and users, thereby fostering a sustainable information ecosystem in the era of AI-generated content.

\backmatter

\bmhead{Competing interests}
The authors declare no competing interests.
\bmhead{Data availability}
All codes developed in this study have been deposited into a publicly available GitHub repository at https://github.com/zgyds0405-hue/Three-party-evolutionary-game-in-misinformation-government.

\bmhead{Ethical approval}
Ethical approval was not required as the study did not involve human participants.
\bmhead{Informed consent}
Informed consent was not
required as the study did not involve human participants.
\bmhead{Author contributions }
Qin Li: Idea formulation, model conception, and theoretical framework
development; Gui Zhang: Simulation, parameter tuning, data analysis, drafting of the original manuscript, and partial contribution to methodological design; Minyu Feng: Paper supervision, conceptual guidance, manuscript revision, editing, final approval, and funding acquisition;  Matja{\v z} Perc: Methodological refinement, editing, and funding acquisition; Attila Szolnoki: Methodological refinement, editing, and funding acquisition.
\bibliography{sn-bibliography}
\end{document}